\documentclass[letterpaper,twocolumn,prl,aps,superscriptaddress,amsmath,amssymb,floatfix]{revtex4-2}
\usepackage{mathptmx}
\usepackage{color}
\usepackage{amsmath}
\usepackage{amssymb}
\usepackage{graphicx}
\usepackage{esint}
\usepackage{siunitx}
\usepackage[T1]{fontenc}
\usepackage[unicode=true,
 bookmarks=true,bookmarksnumbered=false,bookmarksopen=false,
 breaklinks=false,pdfborder={0 0 1},backref=false,colorlinks=true]
 {hyperref}
\hypersetup{
 linkcolor=magenta,urlcolor=blue,citecolor=blue,pdfstartview={FitH},hyperfootnotes=false}

\makeatletter

\usepackage{textcomp}
\usepackage{epstopdf}

\usepackage{amsfonts}

\pdfpageheight\paperheight
\pdfpagewidth\paperwidth

\usepackage{xcolor}\usepackage{soul}
\definecolor{blue}{rgb}{0,0,1}
\definecolor{red}{rgb}{1,0,0}
\definecolor{green}{rgb}{0,1,0}

\usepackage{soul}

\makeatother

\begin{document}

\title{Multifunctional metalens for trapping and characterizing single atoms}

\author{Guang-Jie Chen}
\thanks{These authors contributed equally.}
\affiliation{CAS Key Laboratory of Quantum Information, University of Science
and Technology of China, Hefei 230026, China}
\affiliation{Anhui Province Key Laboratory of Quantum Network, University of Science and Technology of China, Hefei 230026, China}

\author{Dong Zhao}
\thanks{These authors contributed equally.}
\affiliation{Department of optics and optics engineering, University of Science and Technology of China, Hefei, Anhui 230026, China}

\author{Zhu-Bo Wang}
\thanks{These authors contributed equally.}
\affiliation{CAS Key Laboratory of Quantum Information, University of Science
and Technology of China, Hefei 230026, China}
\affiliation{Anhui Province Key Laboratory of Quantum Network, University of Science and Technology of China, Hefei 230026, China}

\author{Ziqin Li}
\affiliation{Department of optics and optics engineering, University of Science and Technology of China, Hefei, Anhui 230026, China}

\author{Ji-Zhe Zhang}
\affiliation{CAS Key Laboratory of Quantum Information, University of Science
and Technology of China, Hefei 230026, China}
\affiliation{Anhui Province Key Laboratory of Quantum Network, University of Science and Technology of China, Hefei 230026, China}

\author{Liang Chen}
\affiliation{CAS Key Laboratory of Quantum Information, University of Science
and Technology of China, Hefei 230026, China}
\affiliation{Anhui Province Key Laboratory of Quantum Network, University of Science and Technology of China, Hefei 230026, China}

\author{Yan-Lei Zhang}
\affiliation{CAS Key Laboratory of Quantum Information, University of Science
and Technology of China, Hefei 230026, China}
\affiliation{Anhui Province Key Laboratory of Quantum Network, University of Science and Technology of China, Hefei 230026, China}

\author{Xin-Biao Xu}
\affiliation{CAS Key Laboratory of Quantum Information, University of Science
and Technology of China, Hefei 230026, China}
\affiliation{Anhui Province Key Laboratory of Quantum Network, University of Science and Technology of China, Hefei 230026, China}

\author{Ai-Ping Liu}
\affiliation{Institute of quantum information and technology, Nanjing University of Posts and Telecommunications, Nanjing 210003, China}

\author{Chun-Hua Dong}
\email{chunhua@ustc.edu.cn}
\affiliation{CAS Key Laboratory of Quantum Information, University of Science
and Technology of China, Hefei 230026, China}
\affiliation{Anhui Province Key Laboratory of Quantum Network, University of Science and Technology of China, Hefei 230026, China}
\affiliation{CAS Center For Excellence in Quantum Information and Quantum Physics,
University of Science and Technology of China, Hefei, Anhui 230026,
China.}

\author{Guang-Can Guo}
\affiliation{CAS Key Laboratory of Quantum Information, University of Science
and Technology of China, Hefei 230026, China}
\affiliation{Anhui Province Key Laboratory of Quantum Network, University of Science and Technology of China, Hefei 230026, China}
\affiliation{CAS Center For Excellence in Quantum Information and Quantum Physics,
University of Science and Technology of China, Hefei, Anhui 230026,
China.}

\author{Kun Huang}
\email{huangk17@ustc.edu.cn}
\affiliation{Department of optics and optics engineering, University of Science and Technology of China, Hefei, Anhui 230026, China}

\author{Chang-Ling Zou}
\email{clzou321@ustc.edu.cn}
\affiliation{CAS Key Laboratory of Quantum Information, University of Science
and Technology of China, Hefei 230026, China}
\affiliation{Anhui Province Key Laboratory of Quantum Network, University of Science and Technology of China, Hefei 230026, China}
\affiliation{CAS Center For Excellence in Quantum Information and Quantum Physics,
University of Science and Technology of China, Hefei, Anhui 230026,
China.}
\date{\today}

\begin{abstract}
Precise control and manipulation of neutral atoms are essential for quantum technologies but largely dependent on conventional bulky optical setups. Here, we demonstrate a multifunctional metalens that integrates an achromatic lens with large numerical aperture, a quarter-wave plate, and a polarizer for trapping and characterizing single Rubidium atoms. The metalens simultaneously focuses a trapping beam at 852\,nm and collects single-photon fluorescence at 780\,nm. We observe a strong dependence of the trapping lifetime on an external bias magnetic field, suggests a complex interplay between the circularly polarized trapping light and the atom's internal states. Our work showcases the potential of metasurfaces in realizing compact and integrated quantum systems based on cold atoms, opening up new possibilities for studying quantum control and manipulation at the nanoscale.
\end{abstract}

\maketitle

\section{Introduction}
\noindent During the last decade, neutral atom array has emerged as a promising experimental platform for exploring quantum technologies~\cite{Saffman2010,Saffman2016,Henriet2020}, and holds the potential for simulating novel quantum matter~\cite{Browaeys2020,Morgado2021}, realizing quantum information processing~\cite{Bluvstein2024,Xu2024}, and high-precision sensors~\cite{Norcia2019,Bornet2023,Schaffner2024}. The defect-free atom array can be constructed using the atomic tweezer array, i.e., tightly focused dipole traps~\cite{Schlosser2001,Schlosser2002,Andersen2022}, through bottom-up approach~\cite{Endres2016,Barredo2016}. Precise control of spatial optical field distribution is essential in the atom array platform. On the one hand, the manipulation of the external states of atoms requires the laser cooling, the optical dipole trap for trapping and transporting of atoms~\cite{Metcalf1999,Miroshnychenko2006,Dorevic2021,Bluvstein2022}. On the other hand, optical control lasers that near-resonant with atomic transitions are required to change the internal states of atoms, store quantum states, induce the long-range coupling between atoms~\cite{Levine2018,Levine2019}. Additionally, the efficient collection of photons emitted by the atom is crucial for detecting the internal quantum state of the atom or generating the long-distance entanglement between the atoms through the photon~\cite{Leent2021}. However, the complicated structured optical fields for multiple manipulation purposes, including single-atom addressed Rydberg excitation, and near-resonant probe optical fields, are challenging to achieve in practice. For instance, multiple optical components working at different wavelengths are required, including the composite lens for realizing large numerical aperture (NA), polarization controlling, dispersion compensation for dipole laser and atomic fluorescence.

The development of nanophotonics offers new opportunities for atom-based quantum technologies~\cite{Kohnen2011,DaRos2020,Liu2022a,Liu2022b,Ovchinnikov2022,Ropp2023,Wang2022,Zhou2024}. Benefiting from the nanofabrication technologies, conventional optical components can be manufactured on a chip, reducing the volume of optical components and enabling the integration of many components on a single chip, which also simplifies the alignment process among optical devices. It has already been demonstrated that conventional three-dimensional magneto-optical traps and the magnetic coils can be integrated onto chips~\cite{Chen2022}. Beyond the compact and stable replacement of free-space components, nanophotonics also enables a new generation of optical components that are extremely challenging for conventional optics. For instance, metasurface structures provide a very efficient approach for realizing novel functional photonic devices by engineering optical nano-antennas on a chip~\cite{Yu2014,Solntsev2021}. Therefore, it is anticipated that the functional metasurfaces could boost the development of atomic physics and devices. 

Recently, it has been demonstrated that a metasurface could be applied as a multiple direction beamsplitter, which allows the direct generation of multiple laser beams for laser cooling~\cite{Zhu2020,Jin2023}. A metalens could replace the conventional long working-distance and large numerical aperture (NA)  lens. And it has been demonstrated that single atoms can be trapped and their fluorescence signals collected.~\cite{Hsu2022}. Additionally, metasurfaces allow for specific customized functions, such as the realization of tweezer array by a single metasurface with a single input beam~\cite{Huang2023,Huang2024,Zeng2024}. These advancements in nanophotonics and metasurfaces have the potential to revolutionize the atom-based quantum technologies~\cite{Zhang2024}, enabling the development of compact, integrated, and highly functional devices for developing robust and scalable atom array~\cite{Yu2024}.

Here, we realized the single atom trapping in a tweezer and collected the single-photon emissions by a multifunctional metalens. The thin metasurface structure, which has a thickness less than 1 micron, realizes a 3-in-1 function, including a waveplates, a polarizer and a high-NA achromatic lens. Other than merely replacing a lens, the metalens also enables the study of the effect of fictitious magnetic field induced by the circularly polarized dipole trap by demonstrating a magentic field-dependent atom heating in the trap. Therefore, the functional metasurface might revolute the atomic, molecular and optics physics by synthesizing, manipulating and characterizing the quantum matter based on precise spatial-temporal controlling of optical fields, thus holds great potential for exploring novel applications and physics of light-matter interactions.

\begin{figure}[t]
\begin{centering}
\includegraphics[width=1\columnwidth]{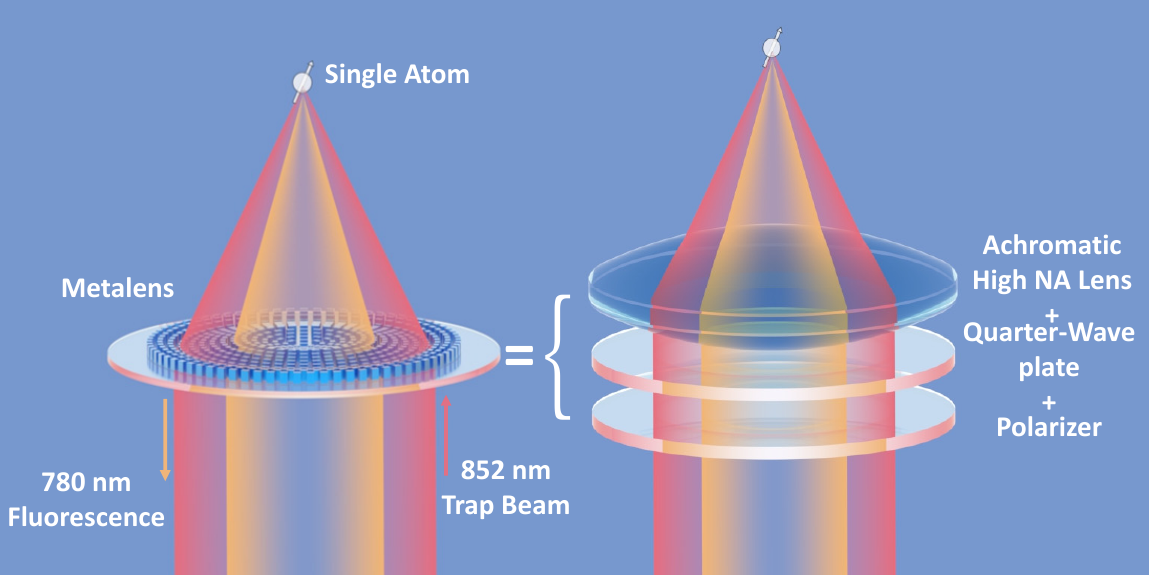}
\par\end{centering}
\caption{\textbf{Schematic illustration of multifunctional metalens for single-atom trapping and characterization.} The metasurface on a thin substrate integrates the functions of an achromatic high-numerical aperture  lens, a quarter-wave plate, and a polarizer. It simultaneously focuses the $852\,\mathrm{nm}$ laser beam to create an optical tweezer for single atoms and collects the $780\,\mathrm{nm}$ fluorescence emitted by the trapped atoms. This compact and integrated device replaces three conventional optical components, simplifying the experimental setup and enabling the study of light-matter interactions at the nanoscale. 
}
\label{Fig1}
\end{figure}

\section{Results}

Figure~\ref{Fig1} schematically illustrates the multifunctional metalens, a metasurface-based device designed for single-atom trapping, manipulation and characterization. The primary objective of this study is to demonstrate the feasibility of realizing a single-atom system using a metasurface chip that integrates multiple essential functions. The metalens must address two key requirements in single-atom experiments: the generation of a tightly focused dipole laser beam (tweezer) for trapping a single atom, and the efficient collection of the fluorescence signals from the trapped atoms. For example, in the case of $^{87}$Rb atoms, due to the time-reversal symmetry of the optical path, it is necessary to combine a normally incident $780\,\mathrm{nm}$ beam and an $852\,\mathrm{nm}$ beam from two independent polarization-maintaining fibers. By focusing them at the same position inside the vacuum chamber, the fluorescence emitted by the atoms trapped by the $852\,\mathrm{nm}$ beam will enter the optical path of the $780\,\mathrm{nm}$ beam and can be efficiently collected. In our design, the metalens simultaneously serves as both the trap and collection optics for the trapped atoms. By carefully engineering the phase profile of the metalens, we ensure that both the $780\,\mathrm{nm}$ beam and the $852\,\mathrm{nm}$ beam are focused at the same point.


Furthermore, we designed the metalens to generate a circularly polarized tweezer that provides a strong fictitious magnetic field~\cite{Tey2008,Chin2017}, which is necessary for efficiently being coupled to the cycling transition of atoms. In particular, it is also of interest in providing a strong field gradient that allows for the strong coupling between the atom's internal state and its external state~\cite{Corwin1999,Miller2002,Thompson2013}. By integrating the quarter-waveplate and polarizer function into our metalens design, we ensure that the single-atom tweezer beam has the desired circular polarization without the need for additional bulky optical components. This 3-in-1 design of the metalens illustrates the key advantages of our approach, which not only reduces the complexity and size of the experimental setup but also opens up new possibilities for studying the interaction between structural optical field and quantum matter at the nanoscale. In the following, we present the design and fabrication of the metalens for $^{87}$Rb atoms, using an $852\,\mathrm{nm}$ dipole trap laser and collecting the $780\,\mathrm{nm}$ fluorescence photons.

\begin{figure*}
\begin{centering}
\includegraphics[width=0.75\textwidth]{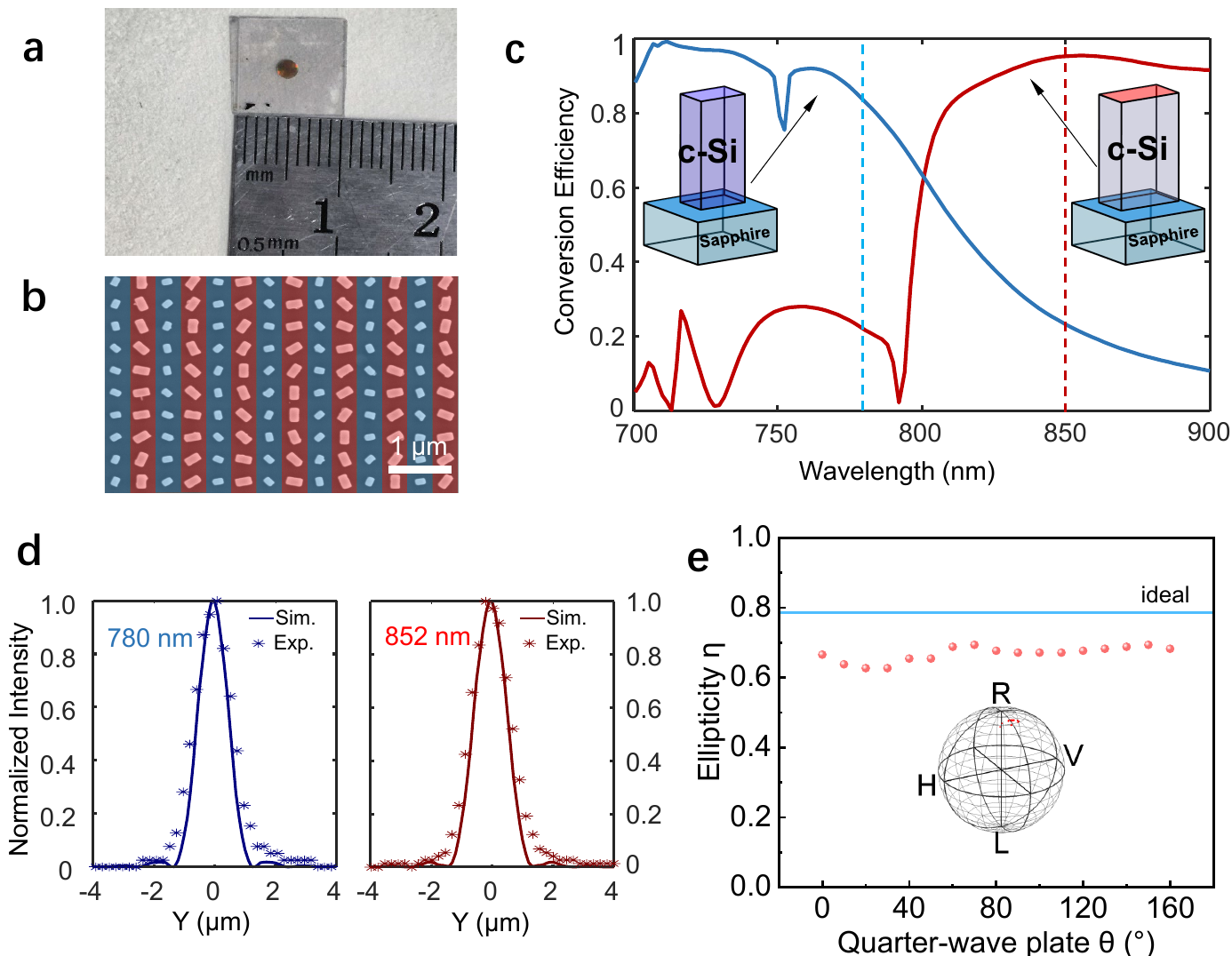}
\par\end{centering}
\caption{\textbf{Design, fabrication, and characterization of the multifunctional metalens.}  (a) Photograph of the metalens chip, consisting of a $10\,\mathrm{mm}\times10\,\mathrm{mm}$ sapphire substrate with a 2\,mm diameter metalens at the center. (b) Scanning electron microscope (SEM) image of the metalens structure, revealing the cross-arranged monocrystalline silicon nanobricks. (c) Simulated wavelength-dependent modulation efficiency of the two types of nanobricks, demonstrating their selective response to 780\,nm and 852\,nm light. (d) Measured and simulated focal spot profiles of the metalens at 780\,nm and 852\,nm wavelengths. The Y direction is one of the radial directions. The metalens achieves achromatic focusing with a spot size of $1.33\,\mathrm{\mu m}$, albeit with some deviation from an ideal Gaussian profile due to fabrication imperfections. (e) Polarization measurement of focused beams using metalens.
}
\label{Fig2}
\end{figure*}

\subsection{Metalens Design and Fabrication}
Figure~\ref{Fig2}(a) presents our metalens chip, which is fabricated on a square sapphire substrate with a length of about $10\,\mathrm{mm}$. The metalens structure is located at the center of the chip, forming a circular area with a diameter of $2\,\mathrm{mm}$. The detailed nanostructures of the metalens are revealed in the scanning electron microscope (SEM) image shown in Fig.~\ref{Fig2}(b). The metalens comprises two types of monocrystalline silicon nanobricks of different sizes, arranged in a cross pattern. We used Finite-Difference Time-Domain (FDTD) to calculate the modulation efficiency of different geometric sizes of the nanobricks at the corresponding wavelength, and then comprehensively considered the efficiency and crosstalk to select the appropriate nanobricks for partition control. The rectangular structure of nanobricks will make the metalens sensitive to the polarization of the incident beam. It will convert the left-handed circularly polarization component in the incident beam into right-handed circular polarization and add a geometric phase. This property makes our metalens function as a right-handed circular polarizer. 

Each nanobrick will modulate the incident light on it, adding a geometric phase to it. The size of this geometric phase is related to the rotation angle of the nanobricks~\cite{Niv2006}, which is twice the rotation angle. We can adjust the rotation angle of these nanobricks to make the geometric phase at different positions different, and finally make the incident beam focused.  In order to achieve achromatism of $780\,\mathrm{nm}$ and $852\,\mathrm{nm}$ incident beam, these nanobricks are specifically designed to modulate incident beam at these two wavelengths. As depicted in Fig.~\ref{Fig2}(c), each type of the nanobricks exhibits a modulation efficiency that is more than three times higher than the other type for its respective dominant wavelength. This difference of modulation efficiency comes from the size of the nanobricks which are $160\times110\,\mathrm{nm}$ for $780\,\mathrm{nm}$ beams and $250\times140\,\mathrm{nm}$ for $852\,\mathrm{nm}$ beams, respectively.  In order to reduce the crosstalk between nanobricks modulating $780\,\mathrm{nm}$ and $852\,\mathrm{nm}$ beam, we specifically adjusted the size of nanobricks modulating $780\,\mathrm{nm}$ so that its highest modulation efficiency is around $720\,\mathrm{nm}$. 

According to equation $\varphi(x, y)=\frac{2 \pi}{\lambda}\left(f-\sqrt{x^{2}+y^{2}-f^{2}}\right)$, we can obtain how much geometric phase the nanobricks of each point on the metalens should give to the incident  beam so that the incident beam can be focused at position $f$ behind the metalens. Here, $f$ is the expected focal length of our metalens, $x$ and $y$ are the distances from each point on the metalens to the center, $\lambda$ is wavelength of the incident beam. By adjusting the angles of the nanobricks, the geometric phase of the incident beams can be precisely controlled while minimizing crosstalk between the two wavelengths. However, because beams of different wavelengths are modulated using two types of nanobricks, the effective focusing efficiency of both wavelengths of  beams drops to about $t=22\%$. This measurement was obtained by filtering the focused beam using a $5\,\mathrm{\mu m}$ aperture. There are three main reasons for the low efficiency: intrinsic absorption of the material, the limitation of partition modulation (not higher than 50\%) and the manufacturing imperfection of the size of the nanobricks. Further study about a high-efficiency dielectric film structure and the design methods that do not use this partition modulation method, but instead build more structural databases and select structures with corresponding phase differences to adjust the parameters of the nanobricks at each position to greatly improve efficiency need to be investigated.

The metalens is fabricated using standard electron-beam lithography (EBL) techniques. First, a $600\,\mathrm{nm}$-thick single-crystal silicon (Si) film is deposited on the sapphire substrate, followed by a 100\,nm-thick electron-beam resist layer (positive, AR-P 6200). 

Subsequently, the dried photoresist is patterned with the mask carrying the structural information of meta-devices using EBL (JEOL, JBX 6300FS) with an accelerating voltage of $100\,\mathrm{kV}$. The patterned photoresist is developed in AR 600-546 for 60 seconds at room temperature, followed by rinsing with flowing deionized water to reduce residue. After dried in an N2 atmosphere, a 10\,nm-thick chromium is deposited on the exposed device to serve as a hard mask. The nanobrick patterns are then transferred into the ultrathin chromium film after lift-off in NMP(N-Methylpyrrolidone) solvent, which is heated in water at $80^\circ \mathrm{C}$ for 5 minutes. Finally, the 600\,nm-thick Si layer without the chromium hard mask is etched by an inductively coupled plasma-reactive ion etching (ICP-RIE) system (Oxford, Plasma Pro System100 ICP380), with an ICP RF power of 1300\,W and a bias power of 50\,W. Once the Si layer is sufficiently etched, the residual chromium mask is removed using a chromium etchant at last, yielding the expected metasurface structure.

The NA of our designed metalens is 0.46. The focusing properties of the metalens are characterized for both $780\,\mathrm{nm}$ and $852\,\mathrm{nm}$ wavelengths, with the results shown in Fig.~\ref{Fig2}(d). We find that the waist of the focused spot is $w_0=1.33\,\mathrm{\mu m}$, with Rayleigh length of $Z_\mathrm{R}=11.68\,\mathrm{\mu m}$. The waist of the focused spot is closer to the focusing ability of a objective with  NA=0.40. However, the Rayleigh length deviate from the expected value of a Gaussian beam focused by a conventional lens with the NA=0.40. For a Gaussian beam with the same waist, the Rayleigh length should be $Z_\mathrm{R}=6.52\,\mathrm{\mu m}$. This discrepancy suggests that the axial confinement of the single-atom tweezer formed by the metalens may be weaker than that of a conventional objective lens with the same NA. It causes our metalens effectively behaves like a lens with an NA of $0.28$ instead of $0.46$ in the axial direction. We attribute this deviation to fabrication imperfections and other factors that cause the metalens to focus the incident light into a beam significantly deviate from the standard Gaussian beam. It is likely that the energy of the incident beam is focused into a small range of different axial positions. The abnormal Rayleigh length of $Z_\mathrm{R}=11.68\,\mathrm{\mu m}$ may be caused by the superposition of the Rayleigh lengths of multiple normally focused Gaussian beams. A evidence of this hypothesis is the side lobes around the main peak in Fig.~\ref{Fig2}(d). Another possible reason is that our metalens was designed with the assumption that the incident beam was a flat-top beam, while in the actual experiment the incident beam was a Gaussian beam with the same diameter as the metalens. Compared with the flat-top beam, the edge energy of the Gaussian beam is lower, and the modulation of the edge structure of the metalens is not fully utilized, which reduces the equivalent $\mathrm{NA}$. In addition, the reduction in edge energy may reduce the energy of the incident beam focused at the center of the focal plane, making the axial energy broadening larger. In subsequent atomic experiments, we find that this defect increases the incident beam power required to trap single atoms and reduce the efficiency for collecting single atoms fluorescence. By measuring the dipole trap depth, we estimate that only about $\zeta\approx33\%$ of the incident beam power contributes to the central peak that forms the effective tweezer for single atoms. This highlights the importance of not only considering the waist size of the focused beam but also paying attention to the Rayleigh length or the side lobes around the main peak, which reflects the beam's energy concentration along the axial direction.

To test the performance of our metalens as a right-handed circular polarizer, we used an objective to re-collimate the focused dipole beam of the metalens and then measured the polarization of this focused beam using a polarization analyzer (SK010PA) 2 meters away, as shown in Fig.~\ref{Fig2}(e). The horizontal axis represents the angle between the fast axis of the quarter-wave plate through which the incident beam passes and the horizontal direction. This angle indicates the varying degrees of circular polarization of the incident beams. The vertical axis is the measured ellipticity $\eta=\cot ^{-1}\left(\sqrt{P_{\mathrm{p}}/{P_{\mathrm{s}}}}\right)$, which is $\pi/4$ when the beam is perfectly right-handed circularly polarized. Our result shows that the $P_{\mathrm{p}}/{P_{\mathrm{s}}}$ of metalens' focused beam remains between 1.3 and 1.9. The beam maintains nearly right-handed circular polarization as designed, regardless of the incident beam's polarization. This observation indicates that our metalens is qualified to function as a right-handed circular polarizer, which will bring a right-handed circular polarization robustness of the dipole beam polarization during single atoms experiments. The difference between this result and the perfect right-handed circular polarizer may come from the birefringence of the sapphire substrate, which needs to be taken into consideration in further research.

\begin{figure*}[t]
\begin{centering}
\includegraphics[width=1.5\columnwidth]{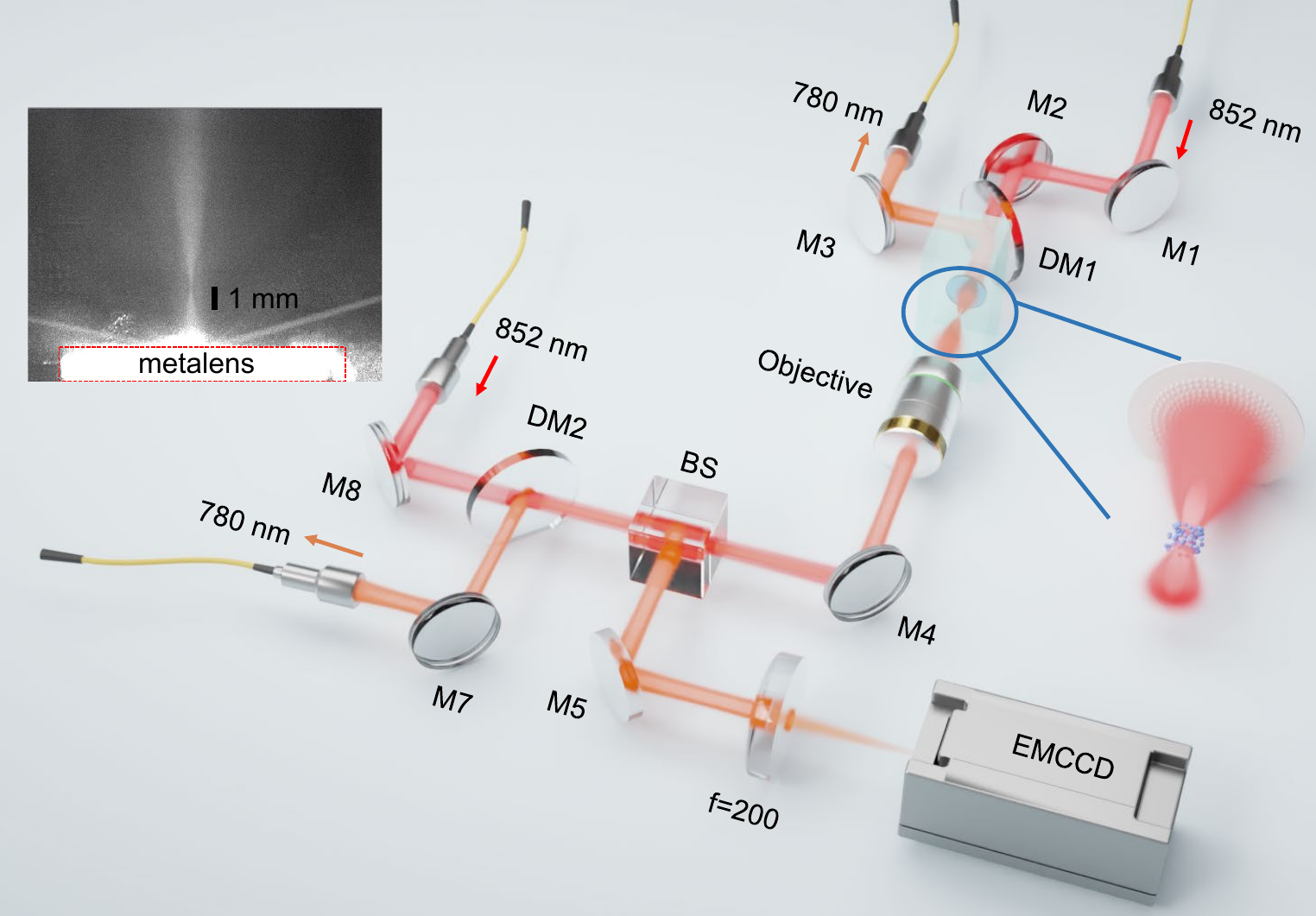}
\par\end{centering}
\caption{\textbf{Experimental setup for single-atom trapping and characterization using the multifunctional metalens.}  Schematic of the optical setup, incorporating the metalens chip at a side of the vacuum cell. A objective with $NA=0.28$ is placed on another side for comparison. The inset is a fluorescence image of the 780\,nm probe beam focused by the metalens, demonstrating a focal distance of 2.6\,mm above the chip surface. Single atoms are loaded into the optical tweezer formed by the 852\,nm beam and their fluorescence will be collected through the same metalens. 
}
\label{Fig3}
\end{figure*}

\subsection{Experimental setup}

The experimental setup, as illustrated in Fig.~\ref{Fig3}, a metalens chip is bonded to the inner wall of the vacuum cell by silicon bonding. The working area of the metalens is as small as 2\,mm, while the remaining part of the chip is transparent to the input cooling beams, allowing the future integration of other on-chip photonic structures on the sapphire chip~\cite{liu2022}. Inside the vacuum cell, a dispenser generates the background gas of $^{87}$Rb atoms, which can be cooled directly by the cooling beam without the Zeeman Slower. Before proceeding with single atom experiments, we first characterized the beam generated by the metalens through the fluorescence of the background Rb atoms gas. The inset of Fig.~\ref{Fig3} shows a side view of the metalens system with an input of $780\,\mathrm{nm}$ laser near-resonant with the D2 transition. The white color indicates the brightness of the fluorescence, and the red dashed frame highlights the saturated bright rectangle caused by the scattering of the substrate. The focused beam generated by the metalens is a vertically propagating beam, that is focused to a very tight waist and then diverges again. This observation confirms that the focus point is about $2.6\,\mathrm{mm}$ above the chip surface, as expected. From this figure, we can also vaguely see the tracks of the beam that are not focused by the metalens but are directly transmitted. Although the total energy of this part of the beam is stronger than that of the focused beam, its light intensity is about $\num{3e-7}$ of the beam waist position. Therefore, they have almost no effect on the metalens trapping atoms. Additionally, two thin beams are deflected to left and right due to the high-order diffraction by the periodic nanostructure of the metalens.

To prepare cold atoms for further single-atom trapping experiments, we adopted the traditional six-beam magneto-optical trap (MOT) configuration~\cite{Metcalf1999}. Four MOT beams are parallel to the chip surface, while the other two beams pass through the metalens chip. All beams have a diameter of only $2\,\mathrm{mm}$ to suppress the scattering by the edges of the chip or by the metalens. A cold Rb atom cloud can be prepared at a distance of $3\,\mathrm{mm}$ from the chip surface, with an atomic temperature of $100\,\mathrm{\mu K}$. It is worth noting that such a metalens configuration is compatible with chip-integrated MOT system~\cite{Chen2022}. Since the chip is transparent except for the center region with a diameter of $2\,\mathrm{mm}$, cold atom could be prepared at the same height as the metalens focal point by aligning and placing a grating chip and a magnetic coil chip outside the cell.

The detailed optical setup for the trapping and characterizing single atoms is illustrated in Fig.~\ref{Fig3}. The core components of the metalens-based single-atom system consist of a dichroic mirrors (DM1) and three mirrors (M1, M2, and M3). Through these components, the normal incident $780\,\mathrm{nm}$ laser and $852\,\mathrm{nm}$ laser from two separate polarization-maintaining fibers could be combined and focused to the same position in the vacuum cell. This process only requires that the two beams are collimated, and does not require any other optical elements to adjust the beams, taking advantage of the achromatic property of our metalens. We also built an independent setup that can also trap and characterizing single atoms using a conventional objective lens (Mitutoyo M Plan APO 10X, $\mathrm{NA}=0.28$, 34\,mm diameter, and 6.6\,cm length). The focal point of this objective overlaps with that of the metalens, which can be realized by either coupling the dipole beam from the metalens to the objective lens or vice versa. The objective-based setup provides a reference system and helps compare the efficiency of the two for collecting fluorescence. Due to the collisional blockade effect, the dipole trap formed by the beam with a radius of only $1.33\,\mathrm{\mu m}$ focused by our metalens will only have two states: 0 atoms and 1 atom~\cite{Schlosser2002}.

\begin{figure*}[!t]
\begin{centering}
\includegraphics[width=1\textwidth]{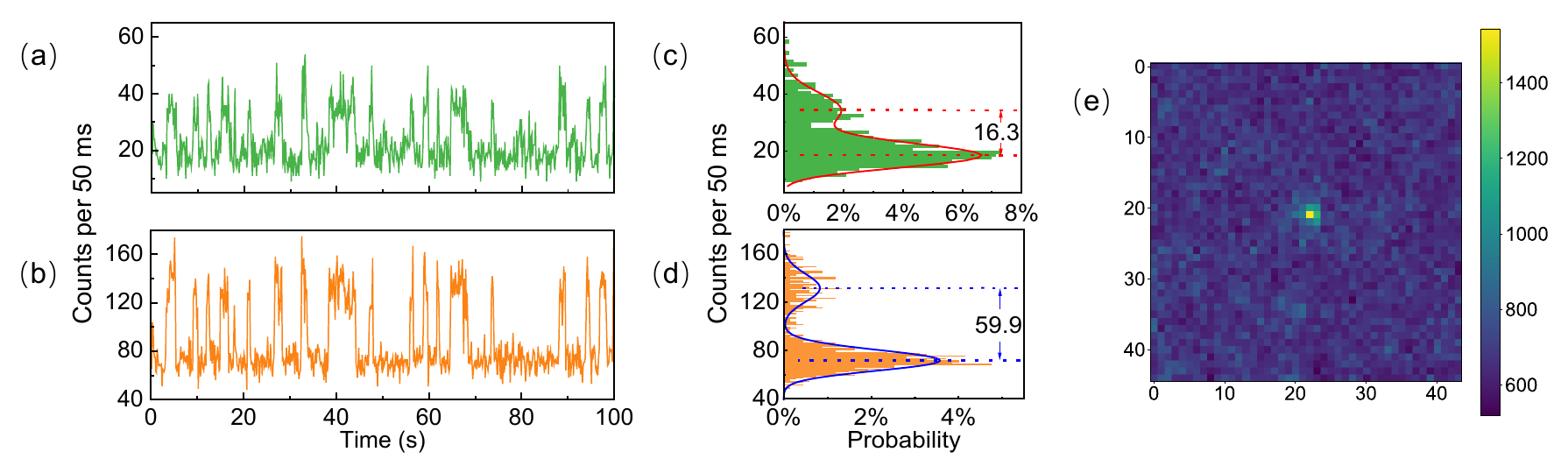}
\par\end{centering}
\caption{\textbf{Single-atom trapping and fluorescence collection using the multifunctional metalens. } (a)-(b) Photon counts from the trapped atom collected by the metalens (green) and the objective lens (orange), Simultaneously. The sudden jumps of the fluorescence in the signal correspond to the loading and loss of individual atoms.  (c)-(d) The histogram statistics of the fluorescence, indicating the mean counts for a single atom is 16.3 and 59.9 for metalens and objective, respectively. (e) Fluorescence signal from a single trapped atom detected by an EMCCD camera in 50\,ms.}
\label{Fig4}
\end{figure*}

\subsection{Single-atom trapping}


By following a sequence of turning on the dipole beam into the metalens, switching the MOT lasers on and off for cold atom cloud preparation, and applying a bias magnetic field $B_{\mathrm{z}}$ along the dipole laser direction, we can trap and probe the single atoms in the tweezer created by the metalens. The duration of each experimental cycle is $4\,\mathrm{s}$ with $2\,\mathrm{s}$ for cold atom cloud preparation and $2\,\mathrm{s}$ for detection. To suppress stray light, the single atoms are probed using a pair of probe beams parallel to the metalens. The probe beams' radius is around $1\,\mathrm{mm}$. The fluorescence scattered by the single atoms captured in the tweezer is collected through our metalens and commercial objectives from opposite directions simultaneously. The collected photons are then guided through the $780\,\mathrm{mm}$ optical path and detect by single-photon counting modules (SPCMs) via single-mode optical fibers. Additionally, we introduced an Electron Multiplying Charge Coupled Device Camera (EMCCD) [Fig.~\ref{Fig3}] to directly imagine the single atoms. 

We repeat the experimental sequences and record the atomic fluorescence when the probe beam is switched on. Typical traces of photon counts per 50\,ms collected by the metalens and objective are plotted in Figs.~\ref{Fig4}(a) and (b), respectively. The dipole beam in the vacuum cell has a power of $P=16.3\,\mathrm{mW}$. The traces exhibit telegraph-like fluorescence signal, with sudden rises and drops in the signal indicating the loading and loss events of single atoms in the tweezer. Figures~\ref{Fig4}(a) and (b) show a representative image of the single atom during our experiment. The signals collected by the EMCCD and the two SPCMs appear and disappear in synchronization on the millisecond scale, confirming that our metalens successfully collects single-atoms fluorescence.

Figures~\ref{Fig4}(c) and (d) are the corresponding histogram of fig.~\ref{Fig4}(a) and (b). It can be found that the intensity of the fluorescence signal collected by metalens is less than $1/3$ of that collected by the objective lens, even though its NA is larger than that of the objective lens. For ideal lenses with $\mathrm{NA}=0.28$ and $0.46$, the corresponding collection efficiency without considering the loss should be $\eta=1.5\%$ and $4.0\%$, respectively. The difference in the collected counts are attributed to the relatively low focusing efficiency of the metalens $t=22\%$ and a factor of beam concentration $\zeta=0.33$. As shown in Fig.~\ref{Fig2}(d), the practical laser beam deviates from a Gaussian beam, with the energy decaying from the center at a slower rate than expected. Compared to the objective lens, which has $t=0.8$ and $\zeta\approx1$, and considering an additional beamsplitter used to separate the fluorescence into two paths for the SPCM and EMCCD, a rough prediction of the ratio of counts $\propto\eta t\zeta$ between the metalens and the objective should be $0.23$, agrees with our experimental results ($0.27$). The fig.~\ref{Fig4}(e) is a typical signal measured by our EMCCD (andor ixon Ultra 888). The color bar means the signal strength in $50\,\mathrm{ms}$.


\begin{figure*}[t]
\begin{centering}
\includegraphics[width=1\textwidth]{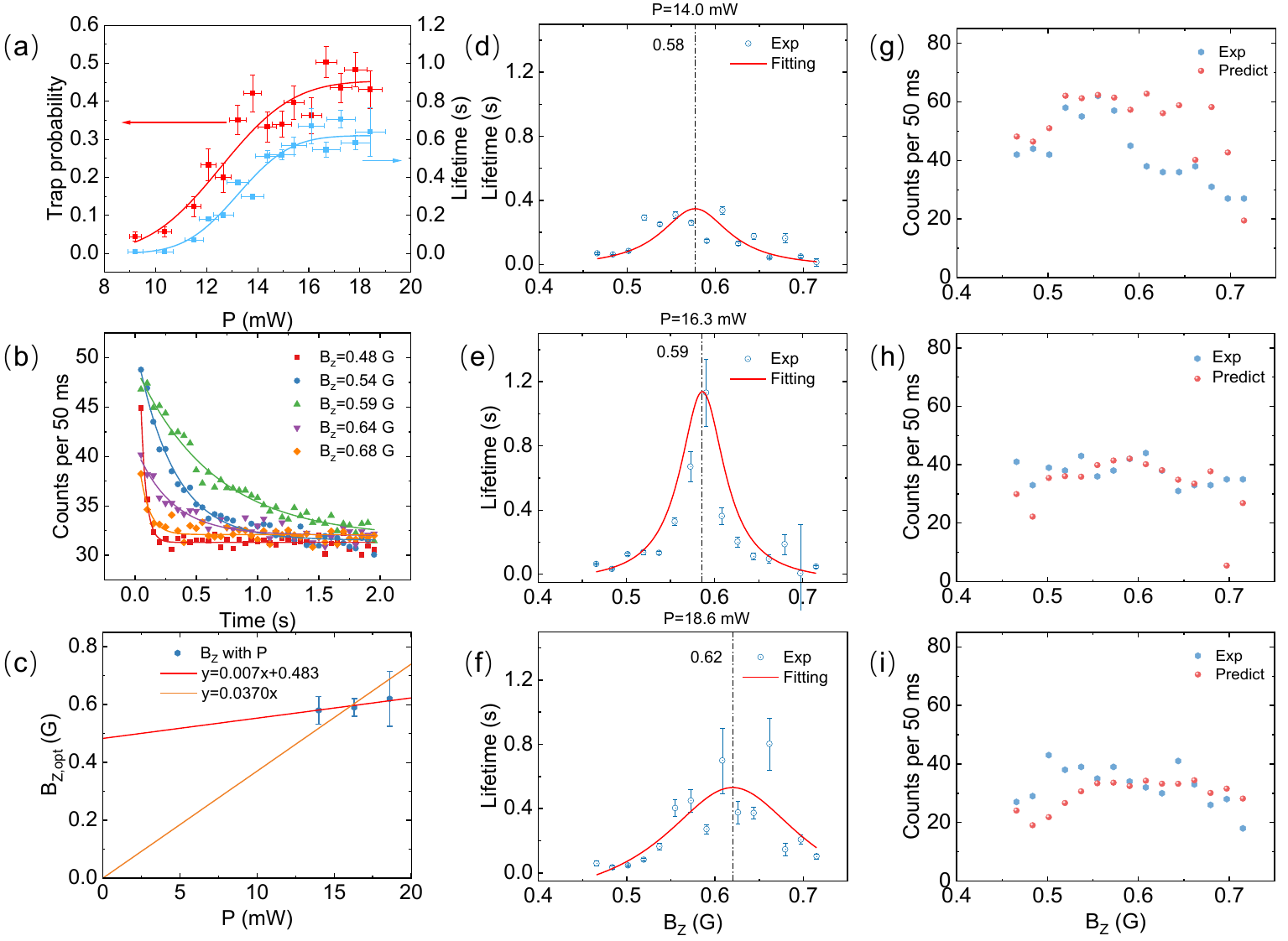}
\par\end{centering}
\caption{\textbf{Magnetic-field dependence of atom dynamics in the circular-polarized metalens tweezer.}
(a) Dependence of the single-atom loading rate and single-atom lifetime on the trapping beam power. The red scattered points are the probability of metalens' dipole trap trapping a single atom, and the blue scattered points are the lifetime of the single atoms in metalens' dipole trap. The red and blue curve is the fit with a error function, respectively. (b) The averaged counts evolution of atomic counts in the tweezer under different bias magnetic field $B_\mathrm{{z}}$. (c) The relation between the optimal $B_\mathrm{z}$ for realizing long-lifetime trapping with a given $P$. (d)-(f) The dependence of trapped atom lifetime for various $B_\mathrm{z}$, with the dipole beam power of $14.0$, $16.3$, and $18.6\,\mathrm{mW}$, respectively. Here, the error bar takes twice the standard deviation. (g)-(i) The corresponding atomic fluorescence step of single atoms. Blue dot: the experimental data, red dots: the fitted counts by taking the lifetime into account. }
\label{Fig5}
\end{figure*}

\subsection{Atom dynamics in a circular-polarized tweezer}

As mentioned earlier, our 3-in-1 metalens can generate a stable circularly-polarized single-atom tweezer, which features a dipole beam intensity dependent fictitious magnetic field ($B_{\mathrm{F}}$) along the propagation direction of the dipole laser. The light-induced fictitious magnetic field can significantly affect the dynamics of atoms within the tweezer when the probe beams exist. Under the influence of probe beams, the single atom scatters photons in random directions, leading to a random walk of its momentum. Consequently, the temperature of the trapped atom increases, and it may escape from the trap, resulting in a limited lifetime due to fluorescence. By employing the polarization-gradient cooling (PGC) configuration, the heating caused by fluorescence can be suppressed, leading to an increased atomic lifetime in the trap, which is eventually limited by the collisions with the vacuum background gas. As studied in previous experiments~\cite{Chin2017}, the $\sigma^{+}-\sigma^{-}$  PGC of a single atom is very sensitive to the polarization ellipticity of the trap beam, and the temperature of trapped atoms increases with the ellipticity, indicates that the PGC is inhibited by the $B_{\mathrm{F}}$. 

To investigate the heating effect related to $B_{\mathrm{F}}$, we carried out experiments by varying the bias magnetic field $B_{\mathrm{z}}$, and the results are summarized in Fig.~\ref{Fig5}. For the results presented in Fig.~\ref{Fig5}, we fixed $B_\mathrm{z}=0.59\,\mathrm{G}$ to optimize the lifetime of atoms under illuminations. By fixing the $B_\mathrm{z}$, we further varied the power of incident beam $P$ and investigated the statistic behavior of the single atom dynamics, as shown in Fig.~\ref{Fig5}(a). The red scattered points (left) are the probability of metalens' dipole trap trapping a single atom, and the blue scattered points (right) are the lifetime of the single atoms in metalens' dipole trap. The red and blue curve is the fit with a error function, respectively~\cite{Schymik2022}. Both the capture probability and the atoms' lifetime conform to the error function, which shows that the dipole trap formed by metalens is no different from that formed by a commercial objective. The trap probability, which represent the probability of finding one atom in the tweezer in each experimental cycle, and the lifetime both increase with the $P$ and saturate at $P>15.9\,\mathrm{mW}$, indicating a trap depth of $0.87\,\mathrm{mK}$ for a $\zeta \approx 0.33$. This behavior is consistent with a conventional linear polarized tweezer. 

Subsequently, We collected the fluorescence signals of single atoms in the dipole trap under different $B_{\mathrm{z}}$ conditions, each signals lasting over one hour. By superimposing and averaging these fluorescence signals, we obtained the average fluorescence signal in one detection cycle. As shown in Fig.~\ref{Fig5}(b), when $B_{\mathrm{z}}$ changes from $0.48\,\mathrm{G}$ to $0.68\,\mathrm{G}$, the decay rate of the average fluorescence signal changes dramatically. When $B_{\mathrm{z}}=0.48\,\mathrm{G}$ and $B_{\mathrm{z}}=0.68\,\mathrm{G}$, the lifetime of a single atom is only about $100\,\mathrm{ms}$. When $B_{\mathrm{z}}=0.59\,\mathrm{G}$, the lifetime of a single atom can reach $1\,\mathrm{s}$. More results at $P=14.0$, $16.3$, and  $18.6\,\mathrm{mW}$ are summarized in Figs.~\ref{Fig5}(d)-(f). We found that there is an optimal $B_\mathrm{z}$ for each $P$, and the lifetime can change by one order of magnitude even when the $B_{\mathrm{z}}$ is changed by only $0.1\,\mathrm{G}$. Such a drastic change cannot be explained by the resonance effect and warrants further investigations in the future. The corresponding atomic fluorescence counts is plotted in Figs.~\ref{Fig5}(g)-(i), showing a $B_{\mathrm{z}}$-dependent count rate. It is found that for $B_{\mathrm{z}}$ value that yield longer lifetime, the fluorescence intensity does not change significantly. It is more likely that the atomic emission rate is insensitive to $B_{\mathrm{z}}$. We predict that the fluorescence intensity reduces due to potential atom escape for short lifetime data points. The predicted fluorescence dependence on lifetime is plotted in Fig.~\ref{Fig5}(g)-(i) as red dots, showing that the trends agree with the experiments. Insensitivity of atom's fluorescence to $B_{\mathrm{z}}$ indicates that the long lifetime cannot be explained by potential dark states of the atoms that inhibit their emission.

We extracted the optimal $B_{\mathrm{z}}$ for different $P$, and the results are plotted in Fig.~\ref{Fig5}(c), indicating an increase in the optimal $B_\mathrm{z}$ with increasing $P$, confirming the increase of the fictitious magnetic field for larger $P$. These effects confirm that the fictitious magnetic field can significantly change the dynamics of atoms in the tweezer. For the model presented in Ref.~\cite{Chin2017}, the longest atom lifetime should be achieved when $B_{\mathrm{z}}+B_{\mathrm{F}}=0$. However, the linear fit of our experimental results violates this relationship as the projected $B_\mathrm{z}\neq0$ for $P=0$, implying the existence of other effects associated with the circular-polarized tweezers, which warrant further investigation. It is also anticipated that the fictitious magnetic field generated by the tightly focused tweezer could induce new physics related to spin-motion coupling, due to the strong gradient of the magnetic field $\sim B_{\mathrm{F}}/w_0$ that cannot be compensated by an external bias field.

\section{Discussion}
The integration of multiple functions onto a single, compact metalens platform offers several key advantages. Our metalens combines beam focusing and polarization control onto a single chip-scale device, reducing the size and complexity of the experimental setup while improving stability and scalability. Furthermore, the metalens can be fabricated using standard CMOS-compatible processes, enabling high-yield, cost-effective production, and facilitating integration with other on-chip components. The ability to design meta-atoms for specific wavelengths allows for the wavelength-selective function, which is potential for further manipulation of atoms involving more laser wavelengths.

To further extend the capabilities of our system, several improvements can be considered. For instance, exploring the use of reflective metasurfaces instead of transmissive ones could potentially increase the overall efficiency of the device and reduce the background noise. Additionally, investigating ways to dynamically control the metasurface properties could enable the realization of reconfigurable trapping potentials, allowing for more flexible manipulation of atom array. 
Another promising direction is the integration of nonlinear optical elements into the metalens design. For example, by incorporating second-harmonic generation or sum-frequency generation, it may be possible to generate entangled photon pairs or perform frequency conversion between the trapping and fluorescence wavelengths. This could enable the realization of hybrid quantum systems that interface atoms with photonic qubits, and the development of long-distance quantum communication networks.

\section{Conclusion}
In conclusion, our work demonstrates the application of a multifunctional metalens for trapping and characterizing single atoms. The device integrates the functions of an achromatic high-NA lens, a quarter-wave plate, and a polarizer, enabling simultaneous focusing of a trapping beam and collection of single-photon fluorescence. The integration of multiple functions into a single metasurface device represents a significant advantage, as it greatly simplifies the experimental setup and reduces the volume of the system. Our results highlight the potential of metasurfaces in realizing compact and integrated atom-based quantum systems, and open up new avenues for studying light-matter interactions at the nanoscale. We anticipate that this approach will find widespread applications in the fields of quantum simulation, quantum information processing, and precision sensing. 

\smallskip{}
\begin{acknowledgments}
We thank Jun-Jie Wang and Hong-Jie Fan for discussions.
This work was funded by the National Key R\&D Program (Grant Nos.~2021YFA1402004 and 2021YFF0603701), and the National Natural Science Foundation of China (Grants No.~U21A20433, No.~U21A6006, and No.~92265108). This work was also supported by the Fundamental Research Funds for the Central Universities and USTC Research Funds of the Double First-Class Initiative. K.H. acknowledges the CAS Project for Young Scientists in Basic Research (Grant No.YSBR-049), the National Natural Science Foundation of China (Grant No. 12134013), the National Key Research and Development Program of China (No. 2022YFB3607300), and the CAS Pioneer Hundred Talents Program. The numerical calculations in this paper have been done on the supercomputing system in the Supercomputing Center of University of Science and Technology of China. This work was partially carried out at the USTC Center for Micro and Nanoscale Research and Fabrication.
\end{acknowledgments}

\bibliographystyle{Zou}
\bibliography{metalens}

\end{document}